\documentclass[preprint,sigconf]{acmart}

\usepackage{booktabs} 
\usepackage{lmodern}
\usepackage{courier}
\usepackage{amssymb}
\usepackage{graphicx}
\usepackage{xspace}
\usepackage{url}
\usepackage[linesnumbered,ruled,noend]{algorithm2e}
\usepackage{rotating}
\usepackage{eucal}
\usepackage{color}
\usepackage{enumerate}
\usepackage{graphicx}
\usepackage{lipsum}
\usepackage{listings}
\usepackage{tikz}
\usepackage{wrapfig}
\usepackage{float}
\usepackage{enumitem}

\SetInd{0.3em}{0.3em}

\newcommand{\s}[1]{\textsf{{#1}}}

\begin{document}

\title{Automated Analysis of Multi-View Software Architectures}

\author{Chih-Hong Cheng}
\affiliation{%
  \institution{fortiss GmbH}
  \city{Munich} 
  \country{Germany}
}
\email{cheng@fortiss.org}

\author{Yassine Hamza}
\affiliation{%
  \institution{Technische Universit{\"a}t  M\"{u}nchen}
  \city{Munich} 
  \country{Germany}
}
\email{yassine.hamza@tum.de}

\author{Harald Ruess}
\affiliation{%
	\institution{fortiss GmbH}
	\city{Munich} 
	\country{Germany}
}
\email{ruess@fortiss.org}


\begin{abstract}
Software architectures usually are comprised of different views for capturing static, runtime, and deployment aspects. What is currently missing, however, are formal validation and verification techniques of multi-view architecture in  very early phases of the software development lifecycle. 
The main contribution of this paper therefore is the construction of a single formal model (in Promela) for certain stylized, and  widely used, multi-view architectures by suitably interpreting and fusing sub-models from different UML diagrams. Possible counter-examples produced by model checking are fed back as test scenarios for debugging the multi-view  architectural model. 	
We have implemented this algorithm as a plug-in for the Enterprise Architect development tool, and successfully used  SPIN model checking for debugging some industrial architectural multi-view models by identifying a number of undesirable corner cases. 
\end{abstract}

%
%




\maketitle


\section{Introduction}


Software architectures usually are comprised of different views for capturing static, runtime, and deployment aspects~\cite{arc42,bass2007software}\@. 
The {static/component view} describes the logical decomposition of the system into building blocks (e.g., packages, components, classes), 
whereas the {runtime view} describes the behavior and interaction of the building blocks as runtime elements in the running system, using diagrams
such as {sequence diagrams}, {activity diagrams}, or {state machines}, 
and the {deployment view} shows how software is assigned to hardware processing and communication elements. 


In the current state-of-the-practice, architectural models are analyzed in early phases in the software development cycle, mainly by means of
manual and resource-intensive review frameworks such as the {\em Architectural Trade-off Analysis Method} (ATAM)~\cite{bass2007software}\@. 
What is currently missing, however, are formal analysis techniques of multi-view architectures for early and automated detection of, say, unwanted behavior due to under-specification. 

In this paper, we therefore reconstruct a single model of a multi-view architecture, which is suitable for formal analysis, by fusing 
sub-models of different views in UML diagrams~\cite{rumbaugh2004unified}, as provided, for example in architectural development tools such  as Enterprise Architect\@. 
Our fusion algorithm proceeds by taking deployment views as skeletons to offer basic communication structure over processes and channels in the actual system. The concrete behavior of each deployed software component --- as documented in the static view --- is captured by run-time views. One notable challenge is to cope with under-specification among views, as dynamic architectural views often only capture certain scenarios but not the complete component behavior and all possible interactions. To this end, \emph{semantic extrapolation} is needed for constructing a model-checkable verification model and we enumerate possible extrapolation strategies.

We have implemented our fusion algorithm as a plug-in for Enterprise Architect (EA)\@.
This plug-in generates verification models in the Promela language, which are used as inputs to the SPIN model checker~\cite{holzmann2004spin}\@.
Counter-examples generated by the model-checkers are used as test cases for debugging the multi-view architectural model.  
We evaluated this EA plug-in in early phases of developing two mission-critical distributed software systems in industrial projects, and successfully identified undesired corner cases due to under-specification in the model. 

\textbf{(Related work)} There is a rich literature on the verification of UML-like diagrams. 
For example, refinement of activity diagrams has been based on LTL model checking~\cite{muram2014automated},
and state machine diagrams have been translated to hierarchical automata as the basis for
model checking~\cite{schafer2001model,niewiadomski2009towards,liu2013formal}\@. 
Moreover, sequence diagrams have a straightforward correspondence to communicating processes and process algebras~\cite{ait_oubelli2011uml,lima2009formal,sieverding2013sequence}. 
Use case diagrams can be checked for consistency or containment by means of viewing them as programs with constraints~\cite{klimek2010formal} or by a translation into activity diagrams~\cite{kosters2001validation}\@. 
Lastly, using  annotations such as UML Marte profile~\cite{gerard2008uml}, one may verify extra-functional properties such as timing~\cite{Louati2015}\@. 
In contrast to these approaches we are analyzing multi-view architectural models, which include static, runtime, and deployment views, being  restricted to a certain stylized use and linking between views. 
We therefore do not address or even try to solve the general multi-view consistency problem 
for UML~\cite{knapp2016multi}\@.




\begin{figure}[t]
	
	\includegraphics[width=0.5\textwidth]{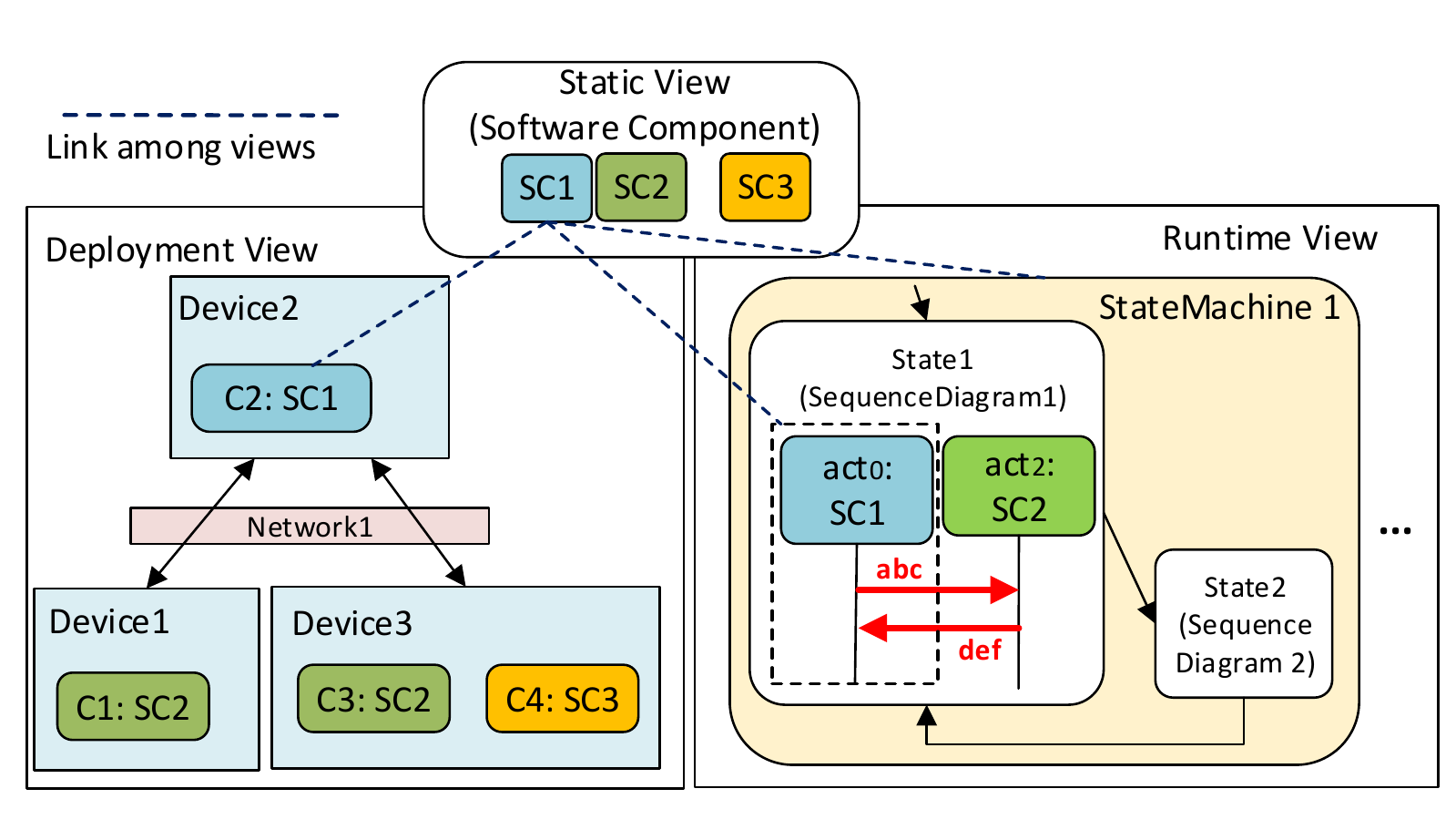}	
	\vspace{-10mm}
	\caption{Example of a stylized multi-view software architecture model.}
	\label{fig:example}
	\vspace{-5mm}	
\end{figure}

\section{Multi-view Software Architecture and View Linking}~\label{sec.multiview}



Using architectural development tools such as Enterprise Architect (EA), the designer may maintain links among multiple views by
creating components in the static view, by building runtime and deployment view using components in the static view,
and by associating each diagram with a component or a sub-structure. 

Figure~\ref{fig:example} illustrates these concepts using a simple architectural example.\footnote{
For the example in Figure~\ref{fig:example}, a model in Enterprise Architect (freely available for model viewing) which maintains such symbolic links can be downloaded at \url{https://www.dropbox.com/s/hg8jiddxh6rs5xs/NFM_Model.eap}. We also refer readers to \url{https://youtu.be/9Mg_2UH5vDM} for a video showing how the link of views are maintained under Enterprise Architect, together with how our prototypical tool automatically generates verification models in Promela form.
}
There are three software components \s{SC1}, \s{SC2}, and \s{SC3} in the static view.
In the deployment view, three devices \s{Device1}, \s{Device2}, \s{Device3} are included in the final deployed system, where for each device, the underlying software components are created (using drag-and-drop in EA) as an \emph{instantiation} of components in the static view. For example, for the \s{Device2} in Figure~\ref{fig:example},  \s{C2} is an instance of the software component \s{SC1} from 
the static view. 

For each software component in the static view there is a state machine or activity diagram in the behavioral view, where each of the states provides behavioral scenarios for different execution modes (for example, normal and error modes)\@. 
Behavior and interaction in each state (or mode) are expressed in terms of scenarios expressed as sequence diagrams. 
In Figure~\ref{fig:example}, for example the behavior of component \s{SC1} is refined to \s{StateMachine1}, where internally, \s{State1} is further refined into \s{SequenceDiagram1}. Notice also that in \s{SequenceDiagram1}, the actor $\s{act}_0$ is surrounded by a dashed component. This is often used in UML modeling as a modeling trick to capture system boundary. 
Such a boundary allows modeling the interaction of multiple instantiations of the same component, 
as commonly seen in fault-tolerant systems where redundancy and distributed voting are applied.

\begin{table}[t]
	\centering
	\resizebox{1.05\columnwidth}{!}{
		
		\begin{tabular}{|p{2cm}|p{3.5cm}|p{4cm}|}				
			\hline
			Elements 	& Meaning &  Corresponding Promela construct  \\ \hline
			$\s{Messages}$\newline $ \{\s{abc}, \s{def}\}$	 & 	Set of message with contents & \texttt{mtype = \{abc, def\};} \\ \hline
			
			$\s{chan1}$
			\newline
			$\in\s{Channels}$  & Synchronous channel	&  \texttt{chan chan1 = [0] of {mtype};} \\ \hline
			
			$\s{chan2[3]}$
			\newline
			$\in\s{Channels}$  & Asynchronous channel named \s{chan2}  with buffer size 3	&  \texttt{chan chan2 = [3] of {mtype};}
			\\ \hline
			Action	$\langle\s{Label: S3}\rangle$	& Program label "S3", move to next action in the process & \texttt{S3:} 		
			\\ \hline
			Action $\langle\s{Goto} \{\s{S3}, \s{S4}\}\rangle$	 & Non-deterministically jump to label  \s{S3} or \s{S4} &
			\texttt{int i;\newline
				select (i : 0..1); \newline
				if \newline
				:: i != 1 -> goto S3\newline
				:: i == 1 -> goto S4\newline
				fi;}  
			\\ \hline
			Action	$\langle\s{chan1 !abc}\rangle$   & Send message \s{abc} to channel \s{chan1}	&  \texttt{chan1 !abc;} 
			\\ \hline
			Action	$\langle\s{chan2 ?def}\rangle$  & 	Receive message \s{def} from channel \s{chan2}	&  \texttt{chan2 ?def;} 
			\\ \hline
		\end{tabular}
	}
	\vspace{2mm}
	\caption{Constructs in verification model and their corresponding formulation in Promela.}
	\label{table.resulting.model}
	\vspace{-5mm}
\end{table}

We are now providing a formal signature for these multi-view architectural concepts;
hereby, $\s{A}.\s{B}$ is used to denote the projection of \s{A} with respect to $\s{B}$\@. 
A \emph{multi-view architectural model} $\s{Arch}$ is a triple $\langle\s{ComponentView}, \s{RuntimeView},\\ \s{DeploymentView}\rangle$\@.  
\s{ComponentView} consists of set of software components where $\s{SC}_i\in \s{ComponentView}$ can 
again be refined to a set of components; for expressing, for example, a ''uses'' structure. For the 
purposes of this paper, such a hierarchical component view can always considered to be in flattened form. 
The \s{DeploymentView} is a pair $\langle \s{Devices}, \s{Network} \rangle$ of sets. 
First, every device $\s{Device}_i \in \s{Devices}$ is a set itself of instantiated software components, and for every
$\s{C}_i \in\s{Device}_i$ is of type $\s{SC}_j$ where $\s{SC}_j \in \s{ComponentView}$\@. 
We use $\s{C}_i.\s{type}$ to denote the typing information. 
Second, pairs of devices $(\s{Device}_i, \s{Device}_j) \in \s{Network}$, where $\s{Device}_i, \s{Device}_j \in \s{Devices}$, 
are interpreted as directed (from left-to-right) edges between devices.
Finally, the \s{RuntimeView} is a quadruple $\langle\s{StateMachines}, \s{SequenceDiagrams}, \s{map}_{SC\rightarrow State}, \s{map}_{State \rightarrow Seq} \rangle$. 
\begin{itemize}[leftmargin=*]
	\item 
	\s{StateMachines} is the set of state machines with each element $\s{SM}_i := (\s{states}_i, s_{0i}, \s{tran}_i)$ having a set of states $\s{states}_i$, an initial state $s_{0i}$ and the set of transitions $s{tran}_i$. We use $\s{SM}_i.\s{s}$ to denote a state~\s{s} in state machine $\s{SM}_i$.
	\item 
	\s{SequenceDiagrams} is the set of sequence diagrams. Again for simplifying formulation, let elements in sequence diagrams be  variable-free, straight-line (i.e., no \texttt{if-else} or \texttt{while}) programs. An element $\s{SeqDiagram} \in \s{SequenceDiagrams}$ is a tuple $(\s{Act}, \s{act}_{0})$, where $\s{Act}$ is the set of actors and $\s{act}_{0}$ is the one that is in the system boundary (cf. \s{act0} in Figure~\ref{fig:example}).  Each actor $\s{act}_i \in \s{Act}$ is a tuple $\langle \s{type}_i, \s{Msg}_i\rangle$ where $\s{type}_i \in \s{ComponentView}$ indicates the typing of the actor by referencing  the element in component view, and $\s{Msg}$ is the finite concatenation of  messages $\s{msg}_{i0}\s{msg}_{i1}\ldots\s{msg}_{ik}$, where forall $j=0,\ldots, k$, 
	$\s{msg}_{ik} \in (\{!,?\} \times \{\s{syn}, \s{asyn}\}\times \Sigma \times \s{Act})$. In  message $\s{msg}_{ik}$,
	$\{!,?\}$ indicates if the message is being sent ($!$) or received ($?$),  $\{\s{syn}, \s{asyn}\}$ indicates synchronous/asynchronous message passing,  $\Sigma$ is used to capture all possible message contents, and the last item is the entity being communicated.
	Consider \s{act0} in Figure~\ref{fig:example}, it is represented as $\langle\s{SC1}, (!,\s{syn},\s{abc},\s{act2})(?,\s{syn},\s{def},\s{act2})\rangle$.
	\item  
	$\s{map}_{SC\rightarrow SM}$ maps an element in \s{ComponentView} to a state machine in \s{StateMachines}.
	For the example in Figure~\ref{fig:example}, \\$\s{map}_{SC\rightarrow SM}(\s{SC1}) = \s{StateMachine1}$.
	\item 
	$\s{map}_{State \rightarrow Seq}$ maps a state in a state machine to zero or one sequence diagram, where if  $\s{map}_{SC\rightarrow State}(\s{SC}_i) = \s{SM}_j$ and if for state $s_j$ in state machine $\s{SM}_j$ we have $\s{map}_{State \rightarrow Seq}(s_j) = \s{SeqDiagram}_k = \langle \s{Act}_k, \s{act}_k = (\s{type}_k, \s{Msg}_k) \rangle$, then $\s{type}_k = \s{SC}_i$. 	For the example in Figure~\ref{fig:example}, \\
	$\s{map}_{State \rightarrow Seq}(\s{StateMachine1.State1}) = \s{SequenceDiagram1}$.
	
\end{itemize}

\begin{algorithm}[t]
	\SetKwInOut{Input}{Input}
	\SetKwInOut{Output}{Output}
	\begin{small}
		\Input{Multi-view architecture model: $\langle\s{ComponentView}, \s{RuntimeView}, \s{DeploymentView}\rangle $}
		\Output{Verification model: $(\s{Processes}, \s{Channels}, \s{Messages})$}
		
		
		\ForEach {$\s{SeqDiagram} \in \s{RuntimeView.SequenceDiagrams}$ }
		{
			\lFor{ $(\s{act}_i = (\s{SC}_i, \s{Msg}_i)) \in \s{SeqDiagram}.\s{Act}$}
			{
				$\s{Messages} := \s{Messages} \cup  \s{Msg}_i $
			}
		}
		
		\ForEach {$(\s{Device}_i, \s{Device}_j) \in \s{DeploymentView.Network}$ }
		{
			$\s{Channels} = \s{Channels} \cup \{\s{chan}_{\s{Device}_i \rightarrow \s{Device}_j}\}$
		}
		
		\ForEach {$\s{Device}_i \in \s{DeploymentView.Device}$ }
		{
			\lFor {$\s{C}_j, \s{C}_k \in \s{Device}_i$ }
			{$\s{Channels} = \s{Channels} \cup \{\s{chan}_{\s{C}_j \rightarrow \s{C}_k}\}$}
		}
		
		\ForEach {$\s{Device}_i \in \s{DeploymentView.Device}$ }
		{
			\ForEach {$\s{C}_j\in \s{Device}_i$ }{
				\textbf{let} $\s{SM}_j = \s{map}_{SC\rightarrow SM}(\s{C}_j.\s{type})$\;
				\textbf{let}	$\s{Pr}_j := \langle \s{Goto} \{\s{s}_0\}\rangle$, where 
				$\s{s}_0$  be the initial state of $\s{SM}_j$\;
				\ForEach {State $\s{s}\in \s{SM}_j.\s{states} $ }{
					
					$\s{Pr}_j := \s{Pr}_j \cdot \langle \s{Label:}\; \s{s}\rangle$ \;	
					
					\textbf{let} $(\s{SeqDiagram}_j = (\s{Act}_j, \s{act}_{0j})) := \s{map}_{State\rightarrow Seq}(\s{s})$\;
					\ForEach {message $(\kappa,\s{syn}, \sigma, \s{act}') \in \s{act}_{0j}.\s{Msg}$, $\kappa \in \{``!", ``?"\}$ }{
						
						\If{$\kappa \;=\; ``!"$ }{
							
							$\s{Pr}_j := \s{Pr}_j \cdot \langle\s{chan}_{\s{C}_j \rightarrow \s{C}_k} !\sigma\rangle$, where $\s{C}_k \in \s{Device}_i$ s.t. $\s{C}_k.\s{type} = \s{act}'.\s{type}$\;
							
							$\s{Pr}_j := \s{Pr}_j \cdot \langle\s{chan}_{\s{Device}_i \rightarrow \s{Device}_k} !\sigma\rangle$, where $\s{C}_k \in \s{Device}_k$ s.t. $i\neq j$ and $\s{C}_k.\s{type} = \s{act}'.\s{type}$\;	
						}\Else{
						
						$\s{Pr}_j := \s{Pr}_j \cdot \langle\s{chan}_{\s{C}_k \rightarrow \s{C}_j  } ?\sigma\rangle$, where $\s{C}_k \in \s{Device}_i$ s.t. $\s{C}_k.\s{type} = \s{act}'.\s{type}$\;
						
						$\s{Pr}_j := \s{Pr}_j \cdot \langle\s{chan}_{\s{Device}_k \rightarrow \s{Device}_i} ?\sigma\rangle$, where $\s{C}_k \in \s{Device}_k$ s.t. $i\neq j$ and $\s{C}_k.\s{type} = \s{act}'.\s{type}$
					}

				}

				\tcc{Jump to successor in state-machine diagram.}
				$\s{Pr}_j := \s{Pr}_j \cdot \langle \s{Goto} \{\;\s{s}'\; |\;\s{s}'\in \s{SM}_j.\s{tran}(\s{s})\;\}\rangle$\;
			}
			
			$			\s{Processes} := \s{Processes} \cup \{\s{Pr}_j\}$
		}
	}
	
\end{small}

\caption{View fusing algorithm}
\label{algorithm.view.fusing}

\end{algorithm}

\section{Multi-View Fusion}~\label{sec.algorithm}

Based on signatures for multi-view architectural models as defined above, we are now describing the process of
providing a behavioral semantics  based on fusing multiple views. 
A \emph{verification model} is a triple $(\s{Messages}, \s{Channels},  \s{Processes})$, where \s{Messages} 
is the set of messages, \s{Channels} is the set of (synchronous or asynchronous-with-fixed-buffer) channels, 
and \s{Processes} is a set of processes. 
Hereby, each process \s{Process} is a sequence of atomic \emph{actions}, including labels, non-deterministic goto primitives, and message send/receive. The semantics of a verification model is based on Promela~\cite{holzmann2004spin}\@. For the purpose of reference, however, we are listing some correspondence of constructs in the architectural model and corresponding verification models in Table~\ref{table.resulting.model}\@.

Now,  the workflow presented in Algorithm~\ref{algorithm.view.fusing}  translates a multi-view architecture into a formal verification model. 
For ease of explanation assume all message passing to be synchronous for now\@.  
Lines~1 and~2 in Algorithm~\ref{algorithm.view.fusing} collect all messages by scanning all actors in the given sequence diagrams. 
Next, lines~3 and~4 define device-to-device channels by scanning through the given network element, 
and lines~5 and~6 define point-to-point channels within a device.  
Lines~7 to~11 start instantiating processes for every deployed software component in the deployment view, 
where the process starts by moving to the initial state (line~10)\@. 
The \s{for}-loop in Line~11 traverses through the state-machine diagram, establishes a label for entry (line~12), and creates outgoing transitions to successor states (line~21)\@. 
Internally, the algorithm jumps to the corresponding sequence diagram (line~13), 
and tries to parse each message being sent or received (line~14) into the corresponding channel (line~16-20), 
where, by probing the deployment view, messages are communicated in the internal channel
if the source and destination components are located in the same device  (line 16, 19). 
Otherwise, intra-device channels are used for communication (line~17, 20). 
Notice that the algorithm simply communicates with all the components having the same type, provided that 
they are supported by the communication architecture in the deployment view. 
This provides the basis for the so-called \emph{extrapolation} in standard UML semantics, as discussed below.

\begin{figure}[t]
	\centering
	\begin{footnotesize}
		\begin{verbatim}
		1    mtype = { abc, def };                   // By line 1-2   
		...
		2    chan Network1_Device2toDevice1Channel = [0] of {mtype}; 
		3    chan Network1_Device2toDevice3Channel = [0] of {mtype};
		4    chan Network1_Device1toDevice2Channel = [0] of {mtype};
		5    chan Network1_Device3toDevice2Channel = [0] of {mtype};
		...
		6   active proctype Device2_C2(){         // By line 7,8,22 
		7     /* Jump to initial state*/
		8     goto State1;                            // By line 10
		9     State1:                                 // By line 12 
		10      /* Contents from sequence diagram */                 
		11      Network1_Device2toDevice1Channel!abc; // By line 17
		12      Network1_Device2toDevice3Channel!abc;                
		13      Network1_Device1toDevice2Channel?def; // By line 20
		14      Network1_Device3toDevice2Channel?def;                
		15      /* Implement the transition*/
		16      goto State2;                          // By line 21
		17    State2:
		18      /* ... (details omitted) ... */ 
		19      goto State1;
		20  }
		\end{verbatim}
	\end{footnotesize}
		\vspace{-3mm}
	\caption{Verification model in Promela form, by running Algorithm~\ref{algorithm.view.fusing} over the example in Figure~\ref{fig:example}.} 	
	\label{fig:result}
	\vspace{-5mm}
\end{figure}

For the example in Figure~\ref{algorithm.view.fusing},  we use the generated verification model in Figure~\ref{fig:result} to explain the concept, where comments in Figure~\ref{fig:result} indicates corresponding actions done in Algorithm~\ref{algorithm.view.fusing}\@. 
Notice that the presentation of the translation algorithm is simplified in that it does not support variables, 
branches and loops. These kinds of extensions are straightforward and are also supported in our prototype implementation.



Most interestingly, lines~16-20 in Figure~\ref{algorithm.view.fusing} make various assumptions about the architectural model
under consideration, and \emph{semantic extrapolation} is used to determine choices being made during the translation.
Such a semantic extrapolation, due to lack of proper semantics in (combining) UML and sometimes due to underspecification in modeling, can be explicitly stated and controlled. 
Table~\ref{table.under.specification} enumerates some important cases and corresponding strategies for semantic extrapolation in order to complete translation.




\section{Evaluation and Concluding Remarks}~\label{sec.conclusion}
\vspace{-1mm}

We have implemented a plug-in for the Enterprise Architect development tool based on the presented translation.
We summarize our findings on using this tool in the architectural design and analysis for two industrial developments.

\vspace{-1mm}

\begin{itemize}[leftmargin=*]
	\item The first case study is a modular adaptive automotive runtime environment. 
	Since this platform has been designed to be fault-tolerant, we annotate possible faults in the deployment view, such as power-outage of a device (fail silent) or lost communication messages. Our tool translates these faults annotation by non-deterministically injecting faults into the generated verification model. 
	In one deployment scenario, a counter-example generated by the SPIN model checker demonstrates that the overall system does not function correctly whenever there are certain faults during start-up, thereby preventing consensus to be reached between computing nodes.
	
	\item Our second case study is a control automation architecture based on the concept of micro-services and a cloud platform. Again, test cases as generated from SPIN model checking of the fused Promela model were instrumental in debugging and improving the design at an early phase in the development. 
\end{itemize}

On the other hand, we have also been experiencing a number of "automation surprises" due to implicit assumptions on the architecture and the generated fused model. For example, the fused model does not capture the fact that service handlers may be viewed as a non-terminating \textsf{while}-loop program that can handle various requests using \s{switch} statements, even though (at least) some designers made such an implicit assumption. 
These kinds of automation surprises might be hard to avoid when applying formal analysis to architectural notations with ambiguous semantics.

It would be most interesting to specify some of the encodings presented here also in a theorem proving environment 
such as PVS, and to  experimentally compare the proposed semantic extrapolation of the behavior of
architectural designs with logic- and constraint-based approaches for partially specified systems.

\begin{table}[t]
	\centering
	\resizebox{\columnwidth}{!}{
		
		\begin{tabular}{|p{5.5cm}|p{4cm}|}				
			\hline
			Under-specification scenarios	& Mitigation strategies  \\ \hline
			In the deployment view, allow components within a device  to communicate with each other?	 & 	\underline{Allow} / Disallow / Trigger the designer for actions \\ \hline
			Operation over variables both in a state of a state-machine diagram and in the refinement sequence diagram of that state?	& Variable operations over variables in a state should appear \{\underline{before}, after\} actions in sequence diagram  \\ \hline
			Unclear requirement in communication buffer size, for asyn. communication?	& \underline{Use pre-defined value} / Trigger the designer for actions \\ \hline			
			An actor sends to one entity in the sequence diagram, while multiple receivers exists in the deployment view?	 & 	\underline{Send to all entities} / Send to one randomly selected entity  / Trigger exception \\ \hline
			
		\end{tabular}
	}
	\vspace{2mm}
	\caption{Semantic extrapolation for handling under-specification in diagrams; the underlined items are strategies used in creating the Promela model in Figure~\ref{fig:result}.}
	\label{table.under.specification}
	\vspace{-7mm}
\end{table}

\bibliographystyle{plain}

\end{document}